\documentstyle[12pt]{article}
\begin{document}
\begin{titlepage}
\vspace*{-1cm}
\begin{flushright}
DTP/94/112   \\
UR-1397 \\
ER-40685-845 \\
November 1994 \\
\end{flushright}
\vskip 1.cm
\begin{center}
{\Large\bf
Gluon Radiation in ${\rm t\bar{t}}$ Production \\[3mm]
at the Tevatron ${\rm p\bar{p}}$ Collider}
\vskip 1.cm
{\large Lynne H. Orr}
\vskip .2cm
{\it Department of Physics, University of Rochester \\
Rochester, NY 14627-0171, USA }\\
\vskip   .4cm
{\large  T. Stelzer}
\vskip .2cm
{\it Department of Physics, University of Durham \\
Durham DH1 3LE, England }\\
\vskip   .4cm
and
\vskip .4cm
{\large  W.J. Stirling}
\vskip .2cm
{\it Departments of Physics and Mathematical Sciences, University of Durham \\
Durham DH1 3LE, England }\\
\vskip 1cm
\end{center}
\begin{abstract}
We present a complete calculation of the matrix elements for the
processes $q \bar q, gg \to b W^+ \bar b W^- g$ and $qg \to
b W^+ \bar b W^- q$ which are relevant for the study of events
with an additional jet in $t \bar t$ production at the Tevatron
$p \bar p $ collider. Our calculation  includes  (i) the contributions
from gluons emitted  during the top production and decay stages and the
interference between these, and (ii) the complete set of Feynman diagrams
corresponding to both resonant and non-resonant  top production.
We study the distribution in phase space of the additional parton jet and
make comparisons
with previous studies based on the soft-gluon approximation
and with results from parton-shower Monte Carlo simulations. The implications
for top
mass measurements are briefly discussed.
\end{abstract}
\vfill
\end{titlepage}
\newpage

\section{Introduction}

A significant number of top quark candidate events reported
by the CDF \cite{CDFTOP} and  D0 \cite{D0TOP} collaborations
at the Tevatron $p \bar p$ collider contain an extra hadronic jet,
in addition to those expected from the leading-order processes
$q\bar q, gg \to t \bar t \to b\bar{b} W^+W^-$. Such jets can be produced,
for example, by gluons emitted from the incoming partons, from the top quarks
before or after they decay, or from the $b$ quarks  in the final state.
Events with such `extra' jets are important from both the experimental and
theoretical viewpoints. Experimentally, they can complicate the
identification and measurement of the top quark, for example when
a gluon jet is wrongly identified as a $b$ jet. They can also, at least
in principle, distort the measurement of the top mass when some
of the top quark four-momentum is carried by a jet which is not
identified as one of the decay products.

There have been several recent studies of extra jets in top production.
In Ref.~\cite{OS}, a complete treatment of all the various
contributions was presented in the `soft-gluon' approximation. This work
built on previous studies \cite{BOOK,PREVIOUS1,PREVIOUS2}
of radiation off heavy
unstable objects.  In Ref.~\cite{LAMPE}, the impact of hard-gluon radiation
on top mass reconstruction was investigated, using a stable, on-shell
top quark approximation which factorizes the gluon emission into
$p\bar{p}\rightarrow t\bar{t}g$ and $t\rightarrow W^+bg$ (or
$\bar{t}\rightarrow W^-\bar{b}g$)
contributions. And of course the parton-shower Monte Carlo programs
used in the experimental analyses naturally give rise to events with
extra jets. However these are based on collinear approximations to
matrix elements and, in some cases, may have certain types of gluon emission
missing.

In this paper we present the first
 complete calculation of the exact matrix elements for the
processes $q \bar q, gg \to b W^+ \bar b W^- g$ and $qg \to
b W^+ \bar b W^- q$, including  (i) the contributions
from gluons emitted during the top production and decay stages and the
interference between these, and (ii) the complete set of Feynman diagrams
corresponding to both resonant and non-resonant  top production.
In the same way that the VECBOS \cite{VECBOS} matrix-element-based
program is successfully used to analyze $W,Z +$ jets events, we would expect
our calculation  to provide  the most accurate predictions for those events
in which an additional energetic jet in association with the
usual $t \bar t$ decay products is observed.

The remainder of the paper is organized as follows.
In the next section we describe how the matrix elements are calculated
using the MadGraph program \cite{MADGRAPH}. We discuss how these
matrix elements are interfaced with the phase space generator to produce
cross sections, and how it is possible to
classify the emitted gluons
as originating  at  either  the top  production or  decay stages.
In Section~3 we describe the  set of kinematical cuts which enables
us to define and calculate a $b W^+ \bar b W^- + $ jet cross section that
is relevant to the experimental measurements.  We present distributions
of the jet  $E_T$ and pseudorapidity, and also of the separation between the
jet and the $b$ quarks, an important quantity when attempting to
reconstruct the top momentum. Some illustrative invariant mass distributions
are also presented.
In Section~4 we compare our distributions with those from the
HERWIG \cite{HERWIG} parton-shower Monte Carlo program, one
of the main analysis tools used in the experimental analyses. We make
a careful comparison of the jet distributions obtained in our
matrix-element  (ME) approach with those generated by the parton-shower
(PS) approach, since any differences {\it could} have important implications
for the measurement of the top mass.
Finally, we present our conclusions in Section~5.

\section{Calculation of the cross section}

As mentioned in the Introduction,
previous works have investigated the effects of gluon radiation
beyond leading order in either the `stable top'  \cite{LAMPE} or
 `soft-gluon' approximations \cite{OS,PREVIOUS1,PREVIOUS2}.
   In this study we focus on
$t \bar t$ production in association with an extra  jet which is
identified as such in the experiment.\footnote{In Ref.~\cite{LAMPE} the
emphasis
was on events with the leading-order number of jets, but which contain
 either additional radiation close to the primary jets  or wide-angle
 radiation not identified as an extra jet.}
 We perform the complete $O(\alpha_s^3)$ tree-level matrix element
  calculation, including  top width effects,
radiation off  the top decay products, and all interferences.

In  $p\bar{p}$ collisions there are three relevant parton-level
processes which may contribute to $t\bar{t}$ production with an
additional jet:
\begin{eqnarray}
1.\  &  q\bar{q} \to b\bar{b} W^+W^- g  \; ,  & \nonumber \\
2.\  &  gg \to  b\bar{b} W^+W^- g       \; , & \nonumber \\
3.\  &  g \bar{q} \to b\bar{b} W^+W^- q \; .  & \nonumber
\end{eqnarray}
Subprocesses one and three are related by crossing and consist of 90
Feynman diagrams, while subprocess two consists of 222 Feynman
diagrams.  The complete set of diagrams and the
corresponding helicity-amplitude code were
generated automatically using the MadGraph \cite{MADGRAPH} package.
In practice, the contribution from subprocess three is very small,
the extra jet is almost always a gluon jet (and
for simplicity will be referred to as such in what follows).
Furthermore, the contribution from subprocess two ($gg$ fusion)
is suppressed by the gluon density in the proton.
Therefore, although the full set of diagrams was used to generate the
figures presented in this paper,
 a reasonable approximation
can be obtained
by using the 7 $q \bar{q} \to t\bar{t}$ production diagrams shown
in Fig.~{1}.\footnote{Note that the  $q \bar q$ annihilation
cross section is an order of magnitude larger than the $gg$ fusion
cross section for $t \bar t$ production  with $m_t \sim 174$~GeV
at the Tevatron collider.} We  neglect radiation off the decay products of the
$W$'s; this is equivalent to assuming either that (i) the $W$'s decay
leptonically, or (ii) strict cuts on the mass reconstruction of the $W$ will
largely eliminate  events where the $W$ decays into
three well-separated jets.
In practice this is a reasonable first approximation, but
in principle the analysis could be extended  to include the hadronic
decay of one of the $W$ bosons, and the corresponding radiation off
the  $q \bar{q}'$ decay products.

In Ref.~\cite{OS}, where the production of an extra jet was analyzed in
the soft approximation, it was shown that the gluon emission could
be separated unambiguously  into `production' and `decay' contributions,
together with interferences between them. The former includes
emission off the incoming partons and off the top quark {\it before}
it weak decays. The latter includes emission off the top quark
{\it during} its decay, and off the daughter $b$ quarks.
The interference between these types of emissions is only important
for soft gluons whose energy is comparable to the top width,
$E_g \sim \Gamma_{t}$ \cite{BOOK,PREVIOUS1,PREVIOUS2}.

In the context of our exact calculation, the separation into
production and decay contributions is not completely
unambiguous (nor gauge invariant in general), but a consistent
operational definition
based on partitioning the phase space can be formulated as follows.
Production emission is defined as those regions of phase space for which the
masses of the $W^- + \bar b$ and $W^+ + b$  systems  reconstruct to the top
mass. Decay emission is defined as those regions of phase space for which
either the $W^- + \bar b$ or $W^+ + b$ system requires the inclusion
of  the extra jet to give the top mass.  This is implemented by comparing
the relative sizes of the Breit-Wigner resonances which appear in the
matrix element.
Explicitly, we define
\begin{eqnarray}
S_{\rm prod} &=&\left|\; ((p_{W^+}+p_{b})^2-m_{t}^2+i m_t \Gamma_{t})\times
((p_{W^-}+p_{\bar b})^2-m_{t}^2+i m_t \Gamma_{t})\; \right| \nonumber \\
S_{1} &=& ((p_{W^+}+p_{b})^2-m_{t}^2+i m_t \Gamma_{t})\times
((p_{W^-}+p_{\bar b}+p_{jet})^2-m_{t}^2+i m_t \Gamma_{t}) \nonumber \\
S_{2} &=& ((p_{W^+}+p_{b}+p_{jet})^2-m_{t}^2+i m_t \Gamma_{t})\times
((p_{W^-}+p_{\bar b})^2-m_{t}^2+i m_t \Gamma_{t}) \nonumber \\
S_{\rm dec} &=& \mbox{min}\,(\,\vert S_1\vert ,\,\vert S_2\vert\,) \;.
\end{eqnarray}
An event is then labeled  as production emission
 if $S_{\rm prod} < S_{\rm dec}$, and as decay emission if
$S_{\rm prod} >S_{\rm dec}$.  This definition is gauge invariant
and  can be used
in any region of phase space for any set of cuts. Since the jet $E_T^{\rm min}$
cut will remove the contribution from very soft gluon emission, one  finds
that for most events which pass the cuts either
$S_{\rm prod} \ll S_{\rm dec}$ or $S_{\rm dec} \ll S_{\rm prod}$.  For
such events production emission is well described by diagrams 1--5 in
Fig.~{1}, and decay emission is well described by diagrams
4--7.\footnote{Note that diagrams 4 and 5 can contribute to {\it both}
types of emission, depending on whether the top quark (or $\bar{t}$)
is closer to being on shell before or after it radiates the gluon.}
  This
decomposition of the radiation is the natural generalization of
 the definition
used in Refs.~\cite{OS,PREVIOUS1,PREVIOUS2}
and proves useful in understanding the
distribution of the radiation.

\section{A study of $ b W^+ \bar b W^-  + $ jet production}

\subsection{Definition of the jet cross section}

Our aim is to calculate the cross section for the production
of an extra, identifiable jet in $t \bar t$ production and decay.
We therefore impose the following cuts on the final state
partons (the subscript $j$ refers to the
extra jet only):\footnote{The cuts are applied to both the
$b$ and $\bar b$ quarks.}
\begin{eqnarray}
|\eta_j| , |\eta_b| \> & \leq & \> 2.5 \; ,\nonumber \\
E_{Tj}, E_{Tb} \> & \geq & \> E_T^{\rm min}= 10 \ {\rm GeV} \; .
\label{cuts}
\end{eqnarray}
In addition, since the extra gluon jet must be distinguishable from
the $b$ jets,  we require the gluon
to be separated  (in ($\eta, \phi$) space)
 from the $b$ and $\bar b$:
\begin{equation}
\Delta R_{bj}, \Delta R_{b\bar b}  \> \geq  \> 0.4  \; .
\label{cutsbis}
\end{equation}
The values of the cut parameters are deliberately chosen to
mimic those in the actual experiments. They also serve to
protect the theoretical cross section from the soft and collinear
singularities of the  matrix element.
Other parameters are: $\sqrt{s} = 1.8$~TeV, $m_t = 174$~GeV,
$\Gamma_t = 1.53$~GeV, $m_b = 5.0$~GeV, and $M_W = 80.0$~GeV.
We use MRS(A) parton distributions \cite{MRSA}
with $\Lambda^{(4)}_{\overline{\rm MS}} = 230$~MeV, $\mu = m_t$
so that $\alpha_s  = 0.103$.

With the above cuts, we obtain the total jet cross section
\begin{equation}
\sigma(p \bar p\to b W^+\bar b W^-+\mbox{jet} + X) \> = \> 2.0\ {\rm pb}\; ,
\label{eq:jet}
\end{equation}
with 51\%  and 49\% coming from the production  and decay
contributions respectively.\footnote{These percentages depend
quite sensitively on the chosen cuts.}
This is to be compared to the leading-order cross section
\begin{equation}
\sigma(p \bar p\to b W^+\bar b W^- + X) \> = \> 3.8\ {\rm pb}\; .
\label{eq:nojet}
\end{equation}
We note that the sum of these is not too far
from the exact next-to-leading order
total cross section of $4.9$~pb calculated in Ref.~\cite{MRSA}.
Although it gives an
indication of consistency in the calculations,
 this equivalence should not be taken too seriously. The cross
sections in Eqs.~(\ref{eq:jet}) and (\ref{eq:nojet}) depend on the cuts
applied to the final-state particles, and we have not included
virtual gluon corrections to the lowest-order cross section.
Furthermore, the $O(\alpha_s^3)$  cross section we calculate here contains
what we
might otherwise think of as corrections to two separate processes.
Roughly speaking, part of the cross section in
Eq.~(\ref{eq:jet}) is part of the $O(\alpha_s)$ correction to the
$\Gamma( t \to bW)$ decay width, while the remainder is part of
the $O(\alpha_s)$ `K-factor' correction to the production cross section.
In the limit $\Gamma_t \to 0$ this correspondence can be made exact --
a full discussion can be found in Sec.~2.3 of Ref.~\cite{PREVIOUS1}.

\subsection{Jet distributions}

In this section we study the distribution in phase space of the
extra gluon jet, for the production, decay and total emission contributions,
as was done in the soft-gluon analysis of Ref.~\cite{OS}.
We note that the cuts we use are slightly different
from those used in Ref.~\cite{OS}, and this
largely accounts for the differences in shapes of some of the
distributions.

Fig.~{2} shows the jet $E_{Tj}$ distribution. The production (dot-dashed
histogram) and decay (dotted histogram) contributions are broadly similar
in shape, with a slight tendency for decay gluons to have higher
$E_T$. Also shown,
for comparison, is the $E_T$ distribution of the $b$ and $\bar b$ quarks.
As one would expect, the latter is significantly harder.
More interesting is the
jet pseudorapidity distribution shown in Fig.~{3}. Here we see a
clear difference between the production and decay distributions. The former
is broad, reflecting the importance of initial state radiation, while the
latter is peaked in the central region, reflecting the tendency
of the decay gluons to follow the directions of the $b$ and $\bar b$
quarks. The distribution is quite sensitive to the $\Delta R_{bj}$
separation cut. Decreasing this from its nominal value of 0.4 has little
effect on the production part, but increases the decay part.
This can be inferred from Fig.~{4}, where
the distribution in the jet--$b$ separation itself is shown.
For production gluons, the dominance of initial-state radiation gives
a broad distribution peaked at $\Delta R_{bj} \sim \pi/2$.
For decay gluons, the collinear quasi-singularity when the gluons are
emitted close to the $b$ quarks (it is not a true singularity because
the $b$'s are massive) gives rise to a sharp peak at small
$\Delta R_{bj}$. This figure illustrates the strong dependence
of the relative proportions of production and decay contributions
on the  separation cut.

\subsection{Invariant mass distributions}

There are obvious problems in  constructing the top mass
from its decay products when there are additional jets in the final state.
For example, the definition $m_t^2 = (p_b + p_W)^2$ will give the correct
mass when the extra jet is a production gluon, but will {\it underestimate}
the mass when it is a decay gluon. Conversely, the definition
$m_t^2 = (p_b + p_W + p_j)^2$ will be correct for decay gluons,
but will {\it overestimate} the mass for production gluons. Of course,
this takes no account of the fact that any gross mismeasurement will be
apparent when one compares the reconstructed masses of the $t$ and $\bar t$.
In principle, one can simply ignore permutations of the decay products
which lead to $m_t \neq m_{\bar t}$.

In practice, however, given the experimental uncertainties in measuring the
jet energies
and in reconstructing one of the $W$ bosons from its leptonic decay products,
there is a danger of biasing the $m_t$ measurement in events with
extra jets, for example by adopting a strategy of not including
such jets in the top quark four-momentum.
For this reason, it is interesting to study the distortion of the top
resonance in the presence of extra jets.

Figure~{5} shows the distribution in $m(bW)$, where
$m(bW)^2 = (p_b + p_W)^2$. There are two entries for each event,
corresponding to combining (say) the $W^+$ with both the
$b$ and the $\bar b$. The dashed histogram (production emission
contribution) simply illustrates the smearing of the resonance peak
from choosing the `wrong' $bW$ combination. The dotted histogram
(decay emission contribution) has a significant shoulder on the lower
side of the peak, showing the effect of omitting a gluon which was
part of the top decay.
The slight dip in the distribution below the peak reflects the $E_T$ cut on the
gluon jet.
The net effect (solid histogram) is a distribution
with a
 strong peak at $m_t$ which is sitting on an asymmetric background,
 with a preference for lower mass values as expected.
For purposes of comparison,
 the insert in Fig.~{5}  shows the leading order
`correct combination' $m_t \simeq  m(bW)$  Breit-Wigner distribution
with width $\Gamma_t = 1.53\ {\rm GeV}$.

When the extra jet is included in the mass reconstruction, the tendency
is to  overestimate the true mass.
This is illustrated in Fig.~{6}, which shows the distribution
in $m(bWj)$, where now $m(bWj)^2 = (p_b + p_W+ p_j)^2$. In this case
it is the production gluon emissions which generate a shoulder above
the peak. There is also a similar effect from decay gluons which are emitted
off the `wrong' top quark.

An interesting feature of Figs.~{5} and {6} is the
much stronger broadening effect in the latter, in which
there is a much larger contribution to the cross section {\it outside}
the main peak.  This is a consequence of the fact that there is a single
extra jet in each event and can be understood as follows.  Let us
ignore wrong $Wb$ pairings and consider only correct ones;
the wrong pairings merely contribute smooth backgrounds to both figures.
In each event at least
one of the $Wb$ pairs will reconstruct to $m_t$ (up to finite-width
and interference effects which we can ignore).  Roughly half of
these events will correspond to production emission, in which case the other
$Wb$ pair will also reconstruct to $m_t$.  Hence $3/4$ of correct $Wb$ pairs
contribute to the $m_t$ peak in Fig.~{5}.  Those same three-quarters of
$Wb$ pairs, when combined with the extra gluon jet, will then typically
fall above the $m_t$ peak in Fig.~{6}.  The remaining one quarter, which
fell below the peak in Fig.~{5}, {\it do}
contribute to the $m_t$ peak when combined with the extra jet, as in
Fig.~{6}.

Figs.~{5} and {6} show, then, that in events with extra jets, one cannot
unambiguously reconstruct the top mass either
by systematically excluding or including
the jet momentum in the reconstruction.  As suggested in Ref.~\cite{OS},
however, one might hope to utilize the different characteristics of
production and decay emissions (as illustrated in Figs.~{2}--{4})
to devise a strategy for deciding whether to include the extra jets in
the reconstruction.
For example, since forward jets tend to be mostly from production
(Fig.~{3}), one could decide to
omit forward jets from mass reconstructions.  In the central region,
where both production and decay jets contribute significantly, one might
gain by making assignments using weighting criteria according to,
for example, proximity to the $b$ quarks.
In any such procedure, of course,
proper account must be taken of hadronization and detector resolution effects,
which are beyond the scope of the present study.

\subsection{Forward--backward asymmetry and color structure}

Soft gluons are able to probe the color structure of a hard scattering
process \cite{BOOK}. In Ref.~\cite{OS} the distribution of the soft
gluon jet was shown to be sensitive to the color structure of the
process $q \bar q\to t \bar t \to b \bar b W^+W^-$ \cite{MW} (see also
\cite{BOOK}).
In particular, the $\widehat{qt}$ antenna (or `string')
produces more radiation in the
region between the $t$ and $q$ than, say, between the $t$ and $\bar q$.
In practice, the effect can be observed by comparing the probability
of gluon radiation between the proton and the $b$ quark
with that between the proton  and $\bar b$ quark \cite{OS}.
We are interested here in whether the asymmetry
observed in Ref.~\cite{OS} in the soft gluon approximation
survives the more exact calculation of the present study.\footnote{Note
that the extra $gg$ and $qg$ processes included here but omitted in
Ref.~\cite{OS} tend to dilute the asymmetry.}

Following the same procedure as in Ref.~\cite{OS} (but with the basic
cuts given in
Eqs.~(\ref{cuts}) and (\ref{cutsbis})),
we define a subsample of   $b\bar b W^+ W^- + $ jet events
in which the $b$ and $\bar b$ are separated by at least
$135^{\circ}$ in azimuth.  This tends to select events in which the parent
$t$ and $\bar t$ have similar separation.   We then
preferentially select gluon jets associated
with the $t$ (as opposed to the $\bar t$) by requiring that they lie
within $90^{\circ}$ in azimuth from the $b$ quark. The $\eta_g$
distribution of such jets should then be asymmetric, with more
jets produced at forward rapidities, {\it i.e.}, between the directions
of the $b$ and the incoming $p$.  Figure~{7}(a) shows
that there is indeed a small forward--backward  asymmetry.
But note that
this asymmetry is a feature of `production' emission only --
the `decay' emission gives a symmetric pseudorapidity distribution
(at least in the limit when $E_g \gg \Gamma_t$ so that interference
contributions can be neglected).
One can therefore enhance the asymmetry by increasing the separation
cut $\Delta R_{bj}$, thereby reducing the decay emission contribution.
 Figure~{7}(b) shows the corresponding
$\eta_g$ distribution when the cut is increased to $\Delta R_{bj}= 1.0$.
It should also be possible to optimize the azimuthal angle cuts
to enhance the effect.

\section{Comparison with HERWIG }

 We have already seen that the cross section for the emission of an extra
 jet in $t \bar t$ production has a very rich structure, with the
 two main contributions coming from production and decay emission.
We have given examples of how this relates to simple top mass
reconstruction  scenarios. It is vitally important that the
programs used in the actual experimental analyses, which must of course
take hadronization and detector effects fully into account and are therefore
much more sophisticated than our parton-level calculations, contain
as much of this structure as possible.

It is not our intention here to make an exhaustive comparison with all
the available programs for simulating top production. Instead, we compare
our predictions for the jet distributions with those of the HERWIG Monte
Carlo program (v5.8)  \cite{HERWIG}, which is widely used in collider physics.
This comparison is not at all straightforward, even when hadronization
is switched off, since the  Monte Carlo program can generate $t \bar t $
events with many additional quarks and gluons  in the final state. We
must therefore introduce a simple jet algorithm  for clustering these
partons. Specifically, we draw cones in
$\eta - \phi$ space around the $b$ quarks and around any additional
energetic partons in the final state, and assign all transverse energy
within the cones to the jet. In this way  we obtain a final state
with $b$-jets (which contain a $b$ quark and possibly other partons)
and additional jets originating in energetic, wide-angle quark
and gluon bremsstrahlung.
The default cone size for clustering  is chosen to be $\Delta R = 0.4$.
We then apply the cuts of Eqs.~(\ref{cuts}) and (\ref{cutsbis}) and
 select those events with one and only one
additional jet,\footnote{With the jet definition used here,
most events with jets contain only one.}
 and compare this  $b\bar b W^+ W^- + $ jet sample
(labeled PS for \lq parton shower' in the figures below) with that
generated by our tree-level matrix element calculation (labeled ME).

Fig.~{8} shows the normalized distributions in (a) the jet-$b$ separation
$\Delta R_{bj}$, (b) the jet pseudorapidity $\eta_j$,
(c) the jet $E_T$, and (d) the jet energy $E_j$ in the
parton subprocess center-of-mass frame.
The first two show significant differences in shape. It would
appear that the PS calculation produces
{\it too few jets in the direction of the $b$ quarks}. In fact
one can see in the PS distribution in Fig.~{8}(a)
 a clear separation between the `production' jets, which are widely
separated from the $b$ quarks in general, and the `decay' jets
which prefer to be close to the $b$ quarks. The ME calculation
evidently produces more of the latter and the dip is filled in.
The same effect is seen in the $\eta_j$ distribution, Fig.~{8}(b).
The peaking at $\eta_j = 0$ in our ME calculation is caused by a sizeable
contribution from `decay' gluons produced close to the centrally-produced
$b$ quarks. Interestingly, the energy distributions shown in  Fig.~{8}(c) are
very similar. However, the preference for more centrally produced
jets in the ME calculation produces a harder jet $E_T$ spectrum,
Fig.~{8}(d).

Fig.~{9} shows the $m(bW)$ and $m(bWj)$ distributions for the ME and PS
calculations.\footnote{For purposes of comparison we include only
the `correct' $bW^+$ and $\bar b W^-$ combinations in Fig.~9.}
  The differences simply reflect the
different behaviours already seen in Fig.~{8}.
There are more PS events in the peak at $m_t$ in the $m(bW)$
distribution, and consequently fewer peak events  in the $m(bWj)$
distribution, since the PS calculation has
apparently fewer jets emitted in  the decay process.

As noted above, the PS calculation requires a jet algorithm to
cluster partons into jets.  We have tried varying the cone size away
from its nominal value of $0.4$ to see if we can improve the agreement
between the ME and PS results. We find, however,
that the changes that result are small in comparison with the
discrepancy (in particular for the $\Delta R_{bj}$ distribution)
between the models.

We have no simple explanation as to why the PS calculation appears to
give qualitatively different distributions
 in jet variables than our exact calculation,
except to note that the differences appear to originate  in the
relative number of jets produced in top production and decay.
Traditionally, ME and PS calculations are expected to agree quite well,
except when the final state contains very energetic, widely-spaced
parton jets; see for example the study in Ref.~\cite{COMPARISON}
for $W$ + jets production. For such configurations, the leading-logarithm
approximation inherent in the PS approach is  expected to break down.
However, we note that our cuts as defined in Eqs.~(\ref{cuts})
and (\ref{cutsbis}) are not very stringent, and
therefore {\it not} particularly
biased in favor of such events. In fact, we have checked that
the $\Delta R_{bj}$ calculated using the soft-gluon approximation
of Ref.~\cite{OS} is very similar to the result of the exact (ME)
calculation shown in Fig.~{8}.

\section{Conclusions}

It is important that experimental data analyses are based on predictions that
account for all relevant physical effects.  In the case of the top quark, one
very relevant physical effect is the presence of extra jets in top events.
We have presented the results of the first exact calculation of hadronic
$t\bar{t}$ production and decay in association with an additional jet,
taking into account all Feynman diagrams that can contribute, including
both resonant and non-resonant top production and all interferences.
This extends the work of Ref.~\cite{OS} to gluons of arbitrary energies,
and allows for a more complete and exact analysis.

We have studied the distribution of such extra jets, and showed that the
cross section can be decomposed into emissions associated with top
production and with top decay, according to a gauge-invariant operational
definition.  This decomposition is particularly relevant to top
mass reconstruction, and
our motivation was, in part, to consider the consequences for top mass
measurement of the presence of such extra jets in top events.
We have seen that extra jets can give rise to shoulders outside
the Breit-Wigner
peak in $Wb$ and $Wb+{\rm jet}$ invariant mass distributions, potentially
degrading the resolution for top mass measurements.

We have also considered distributions generated with the
HERWIG parton-shower Monte Carlo program,
which is a major analysis tool for the experiments and which uses a
collinear approximation to simulate emission of gluons.  After defining a jet
clustering algorithm that allowed us to compare our matrix element results
with those of HERWIG, we have found some discrepancies.  These appear to
indicate a  relatively smaller contribution from  decay gluons
generated by HERWIG, but we are unable to
explain the difference in detail.

\section*{\Large\bf Acknowledgements}

\noindent Two of us (TS,WJS) are grateful to the UK PPARC
 for a Post-Doctoral and Senior Fellowship respectively.
Useful discussions with Valery Khoze, Michelangelo Mangano, and Bryan Webber
 are acknowledged.
This work was supported in part by the U.S.\ Department of Energy,
under grant DE-FG02-91ER40685 and by the EU Programme
``Human Capital and Mobility'', Network ``Physics at High Energy
Colliders'', contract CHRX-CT93-0537 (DG 12 COMA).
\goodbreak

\goodbreak

\section*{Figure Captions}
\begin{itemize}

\item [{[1]}] The subset of (7) Feynman diagrams
which dominate the cross section for the production of extra
jets in top production.

\item [{[2]}]  The  $E_T$ distribution  (solid histogram)
of the extra jet
produced in association with  $t \bar t$
in $p \bar p$ collisions at $\sqrt{s} =1.8$~TeV, together with
its decomposition in terms of
production (dot-dashed
histogram) and decay (dotted histogram) emission contributions.
Also shown is the $E_T$ distribution of the $b$ and $\bar b$ quarks.

\item [{[3]}] The extra jet pseudorapidity  ($\eta_j$) distribution
(solid histogram) and its
decomposition in terms of production (dot-dashed
histogram) and decay (dotted histogram) emission contributions.

\item [{[4]}]  The Lego-plot jet-$b$ separation ($\Delta R_{bj}
 = (\Delta\eta_{bj}^2 + \Delta\phi_{bj}^2)^{1/2}$) distribution
(solid histogram) and its decomposition in terms of production (dot-dashed
histogram) and decay (dotted histogram) emission contributions.

\item [{[5]}] The distribution (solid histogram)
in the $Wb$ invariant mass,  $m(bW)^2 = (p_b + p_W)^2$.
Also shown are the distributions corresponding to the
production (dot-dashed
histogram) and decay (dotted histogram) emission contributions.

\item [{[6]}] The distribution (solid histogram)
in the $Wb+{\rm jet}$ invariant mass,  $m(bWj)^2 = (p_b + p_W
+p_j)^2$.
Also shown are the distributions corresponding to the
production (dot-dashed
histogram) and decay (dotted histogram) emission contributions.

\item [{[7]}] The jet pseudorapidity asymmetry distribution
(solid histogram) defined in the text, and its decomposition in terms of
production (dot-dashed
histogram) and decay (dotted histogram) emission contributions,
for (a) $\Delta R_{bj} > 0.4$ and (b) $\Delta R_{bj} > 1.0$.

\item [{[8]}] Distributions in (a) the jet-$b$ separation
$\Delta R_{bj}$, (b) the jet pseudorapidity $\eta_j$,
(c) the jet $E_T$, and (d) the jet energy $E_j$ in the subprocess
center-of-mass frame, for the exact calculation (solid histograms,
labeled ME) and as obtained using the HERWIG parton-shower Monte Carlo
program (dashed histograms, labeled PS).

\item [{[9]}] As in Fig.~{8}, but for the
$Wb$ and $Wb+{\rm jet}$ invariant mass distributions.

\end{itemize}

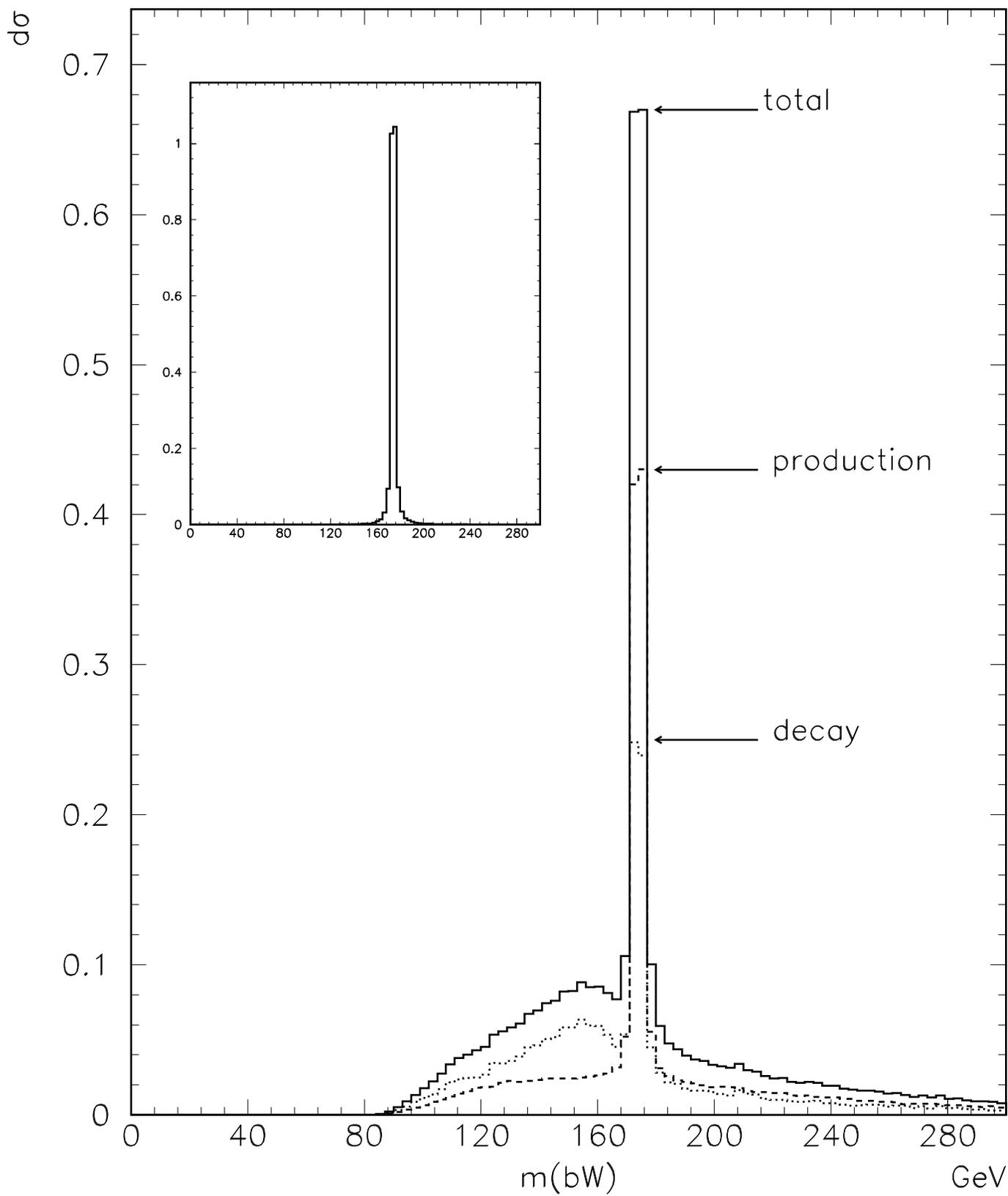
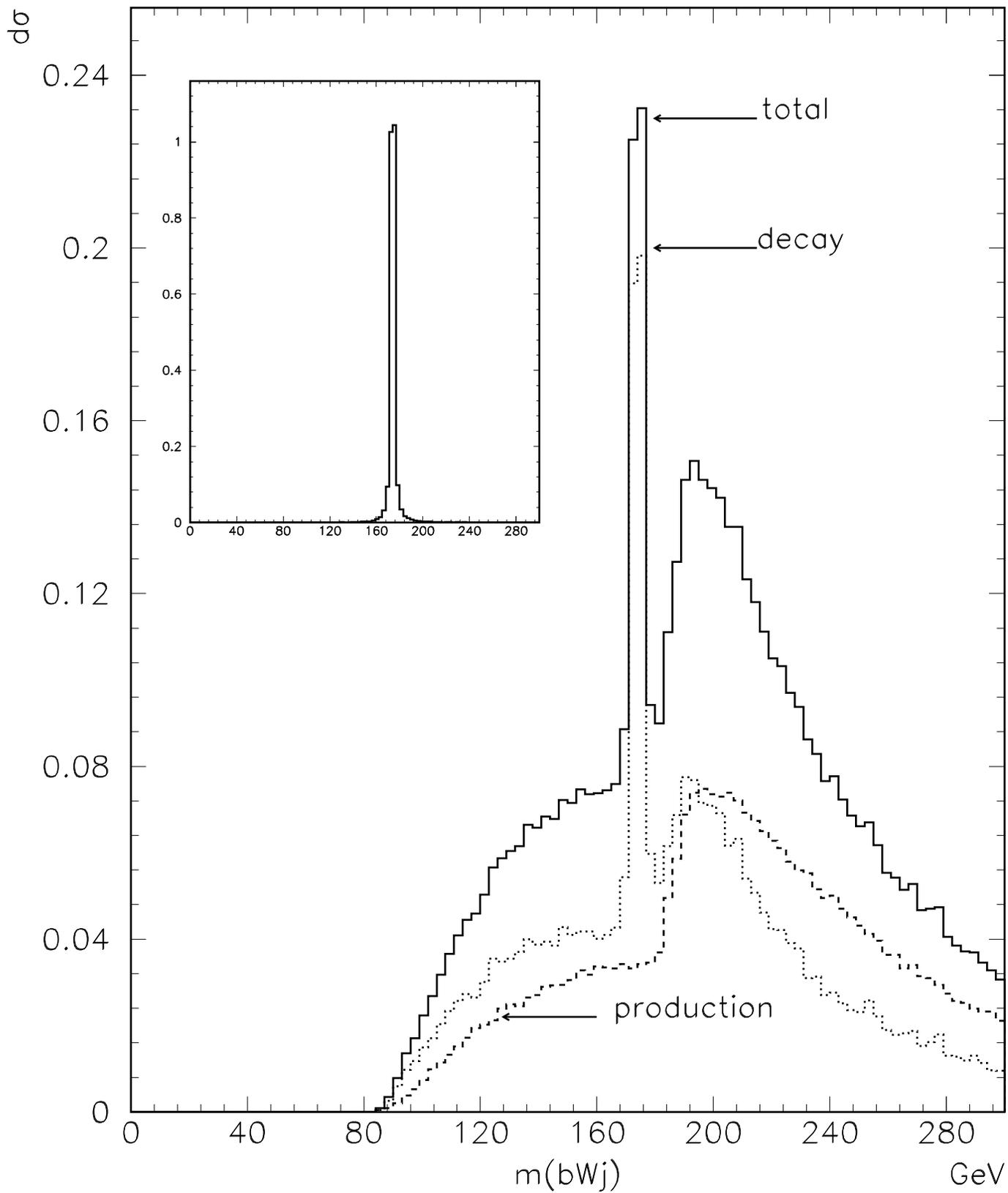
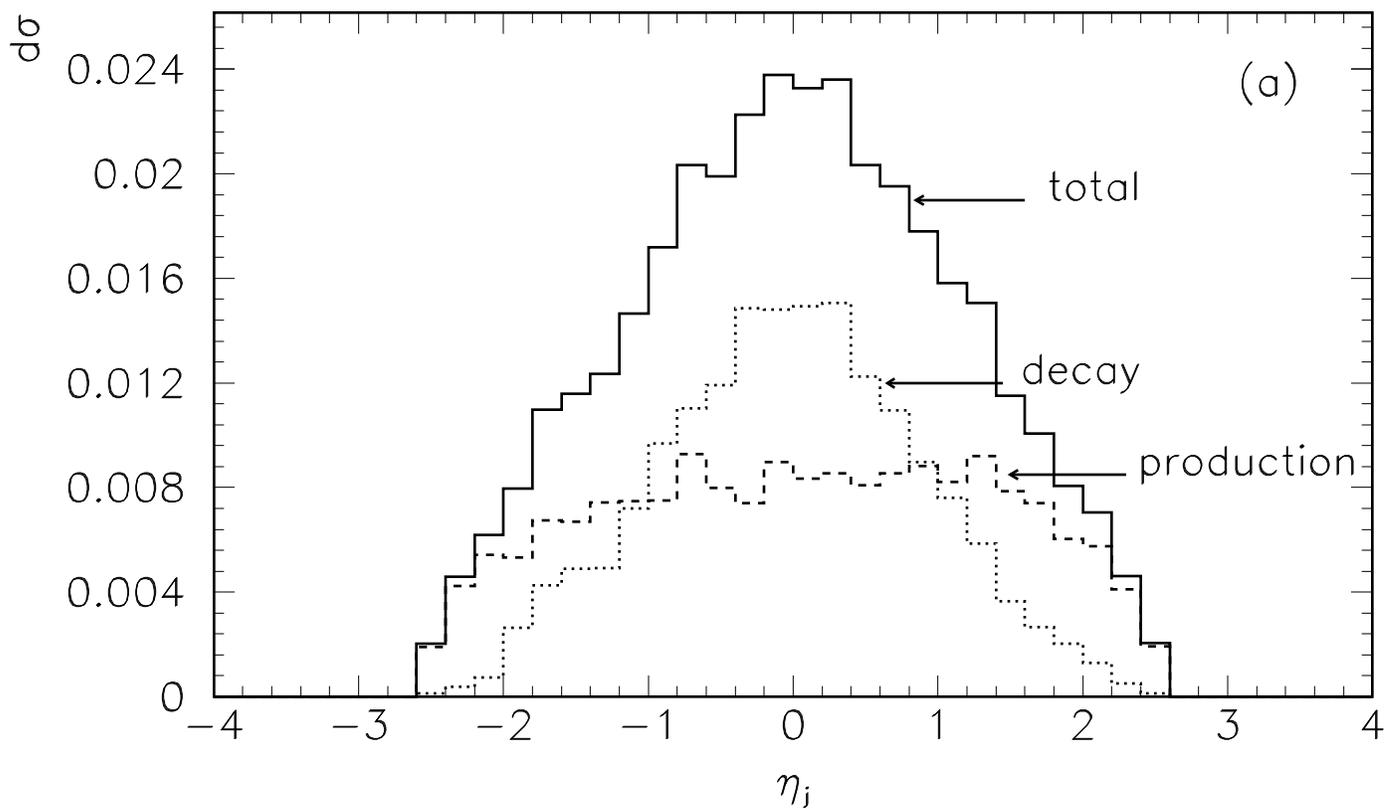
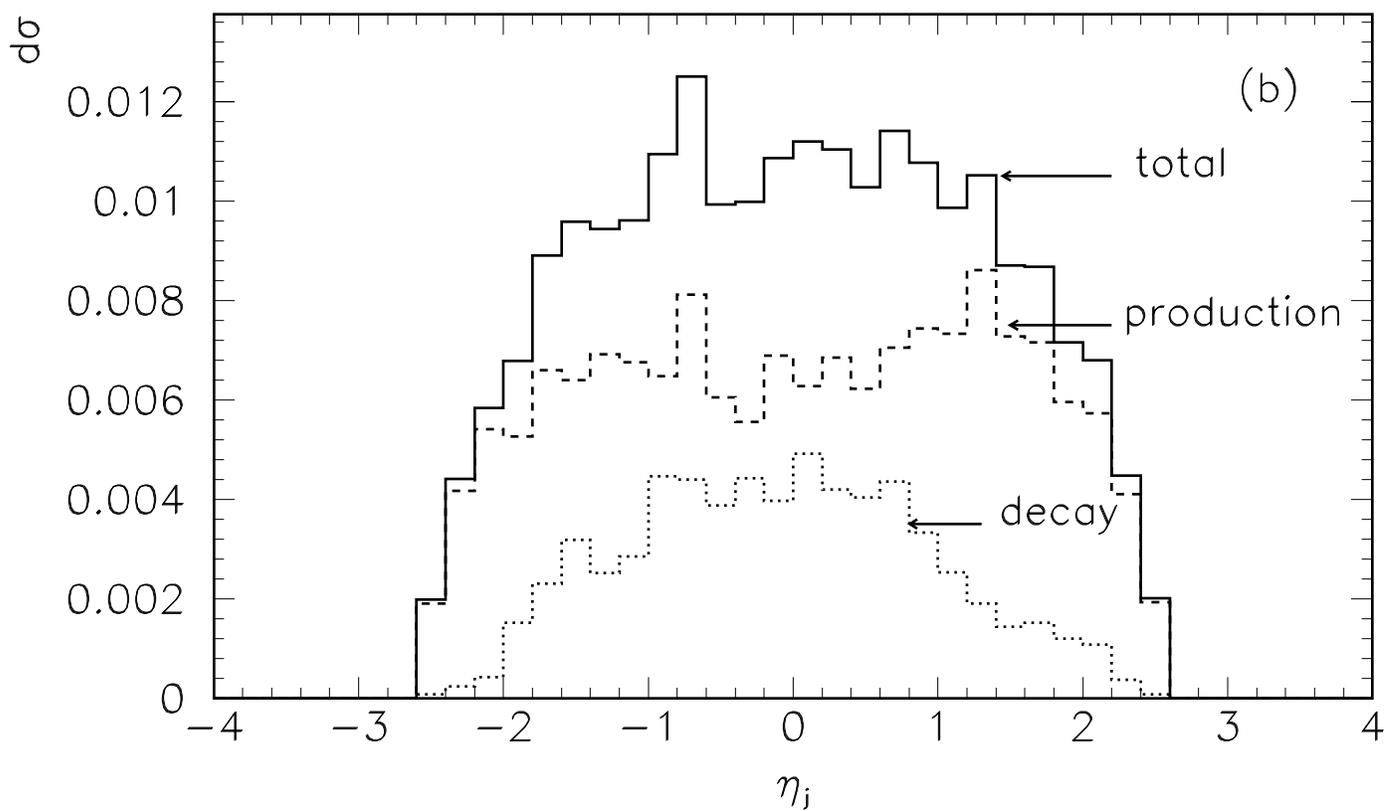
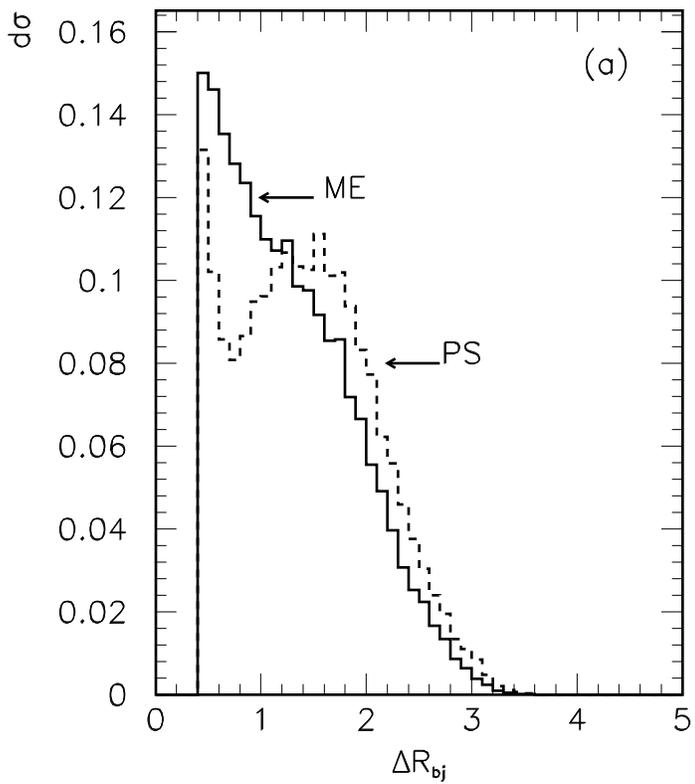
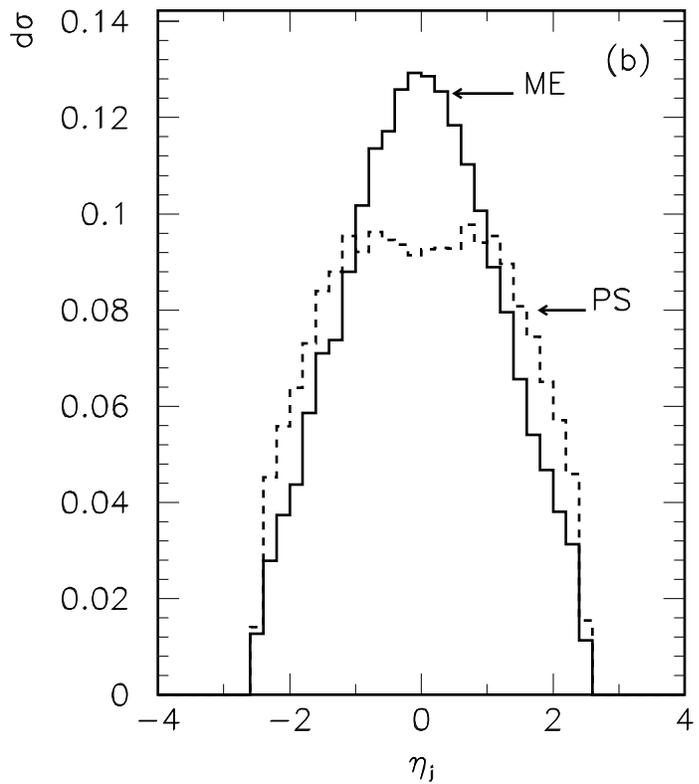
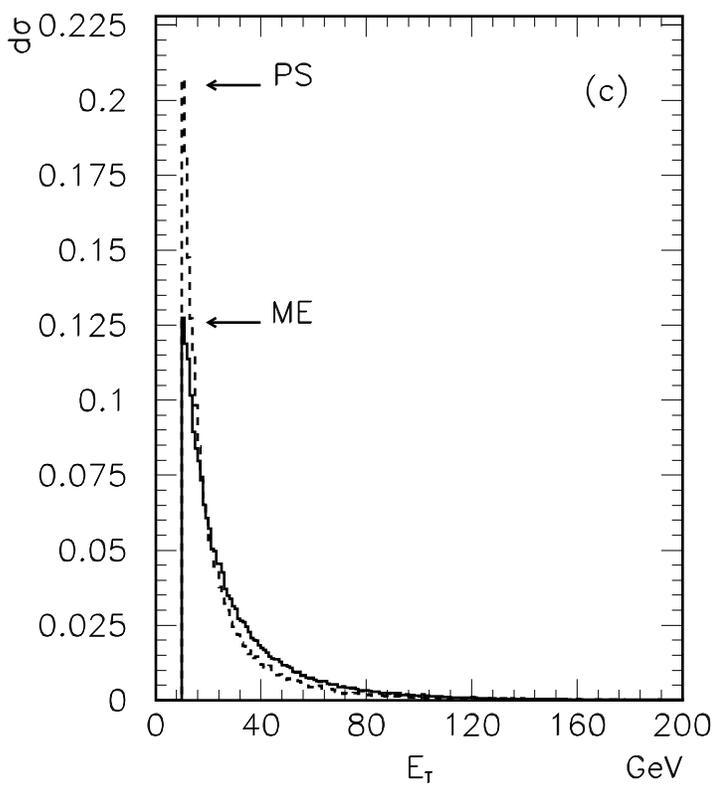
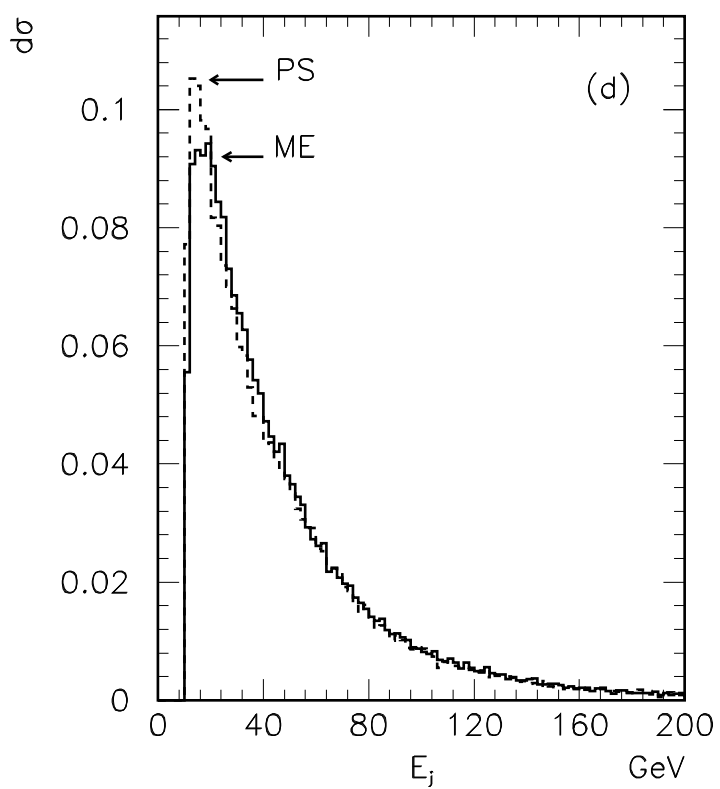
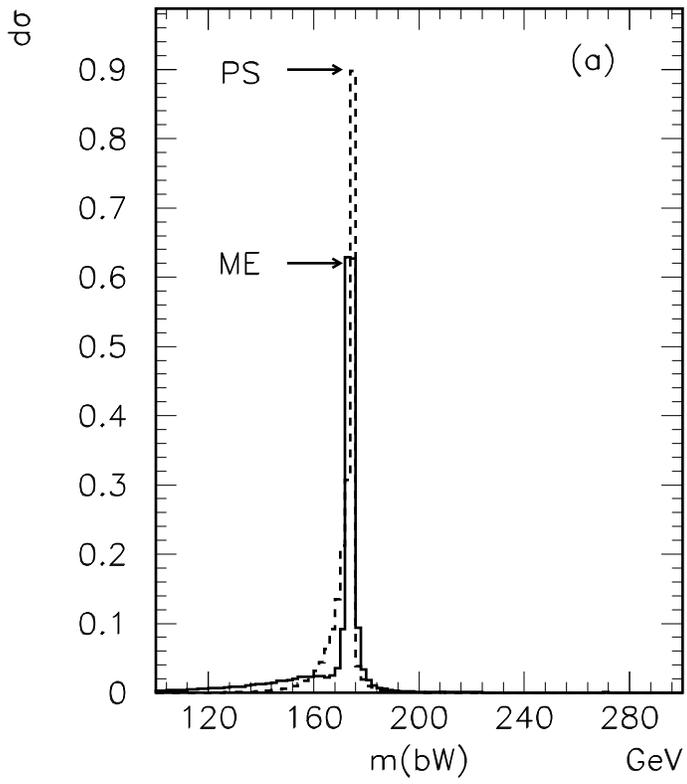
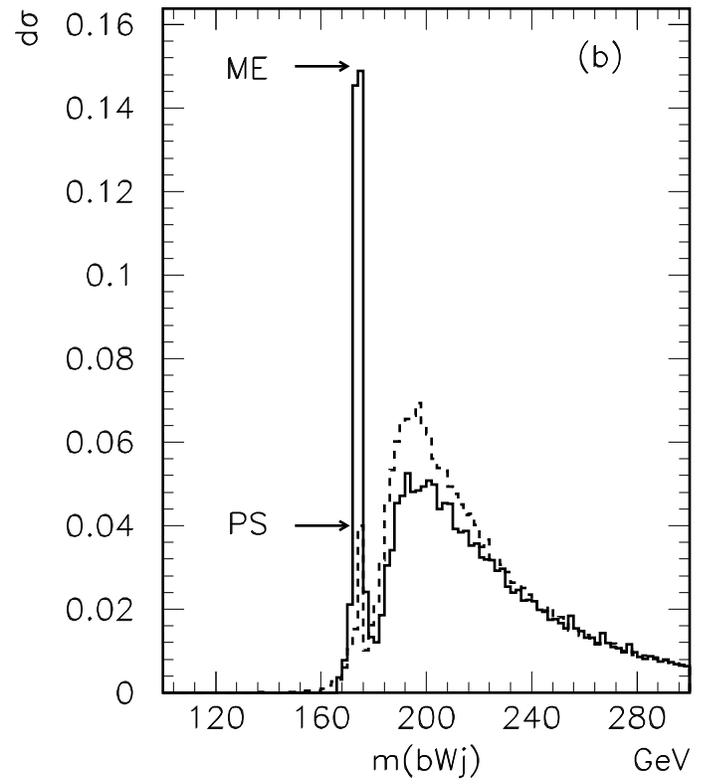
\end{document}